\documentclass[draft]{iopart}
\usepackage{iopams}
\usepackage[dvips]{color}
\usepackage{hyperref}

\begin{document}
\title[Spinor holographic formula]
{Determinant and Weyl anomaly of Dirac operator: a holographic derivation}

\author{R Aros and D E D\'{\i}az}

\address{Universidad Andres Bello,
Facultad de Ciencias Exactas,
Departamento de Ciencias Fisicas,
Republica 220, Santiago, Chile}
\ead{raros,danilodiaz@unab.cl}

\begin{abstract}
We present a holographic formula relating functional determinants: the fermion
determinant in the one-loop effective action of bulk spinors in an
asymptotically locally AdS background, and the determinant of the two-point
function of the dual operator at the conformal boundary. The formula originates
from AdS/CFT heuristics that map a quantum contribution in the bulk partition
function to a subleading large-N contribution in the boundary partition
function. We use this holographic picture to address questions in spectral
theory and conformal geometry. As an instance, we compute the type-A Weyl
anomaly and the determinant of the iterated Dirac operator on round spheres,
express the latter in terms of Barnes' multiple gamma function and gain insight
into a conjecture by B\"ar and Schopka.
\end{abstract}

\section{Introduction}

Ever since its appearance almost fifteen years ago in the form of Maldacena's
conjecture, the AdS/CFT correspondence
\cite{Maldacena:1998re,Gubser:1998bc,Witten:1998qj} has been a successful tool
to address questions concerning strongly-coupled systems. Many developments
depart from the original canonical formulation in pure anti-de Sitter
(AdS) spacetime (mostly still restricted to classical (super)gravity
in the bulk); for instance, finite-temperature effects on the
boundary theory led to consideration of AdS black holes as bulk background
geometries. This novel holographic approach, phenomenological in nature, covers
an increasing amount of physical situations ranging from strongly coupled
quark-gluon plasma and condensed matter systems to cosmological singularities
and black hole physics. At present, this ambitious but conjectural program seems
to succeed at a qualitative level (cf.~\cite{Karch:2011ds}).

In contrast, quantitative and exact results in AdS/CFT correspondence are to be
found in the interplay with mathematics; in for instance conformal geometry and
spectral theory. Geometric roots of AdS/CFT date back to the seminal work of
Fefferman and Graham~\cite{FG85,FG} that addresses conformal geometry on a
compact manifold as geometry at the conformal infinity of space-filling Poincare
metrics. Conversely, AdS/CFT revealed interesting conformal invariants, e.g.
Q-curvature, that arise in the volume renormalization of these Poincare metrics
and triggered new developments in conformal geometry
(cf.~\cite{Gra99,Albin:2005}).

The present contribution will focus precisely on these latter aspects of the
duality, where the foreseeable progress seems modest but solid. We deal with
`holographic formulas' as special entries in the AdS/CFT dictionary, relating
one-loop determinants for bulk fields in asymptotically AdS backgrounds and
determinants of correlation functions of the dual operators at the boundary.
They originate in quantum refinements of the duality where one-loop corrections
in the gravity side are mapped to sub-leading terms in the large-N expansion of the
boundary theory. Interesting effects in for instance thermodynamics
and transport phenomena on the boundary are captured by the holographic
correspondence only after inclusion of quantum one-loop effects in the bulk
(cf.~\cite{Denef:2009kn,BolognesiTong}). A systematic study of bulk {\it
scalars} has led to a holographic formula which has been verified
in certain cases amenable to analytic evaluation; these bulk geometries
include pure and thermal AdS, the BTZ black hole, and other quotients or
orbifolds of AdS~\cite{Diaz:2007an,Diaz:2008hy,Diaz:2008iv,Aros:2009pg}.

Our aim now is to show that also for bulk {\it spinors} an analogous holographic
formula can be established. Explicit computations are performed for the ball
model of hyperbolic space which embrace several scattered results in the
literature. In this case the bulk side and the role of Barnes' gamma function,
already explored in~\cite{Basar:2009rp,Aros:2010ng}, can be further exploited
to get a closed formula for the determinant of the Dirac operator on round
spheres, an interesting result that seems to have escaped notice.
A universal formula for the type-A trace anomaly~\cite{Deser:1993yx} of Dirac operator is as well
obtained in this holographic way. These explicit results contain previous
ones found in relation with proposals for a c-theorem in dimensions other than two;
they include Cardy's a-theorem~\cite{Cardy:1988cwa,Allais:2010qq}, universal terms
in entanglement entropy~\cite{Dowker:2010bu,Myers:2010tj} and
F-theorem~\cite{Klebanov:2011gs}.

We start in section 2 with a review of bulk spinors in AdS/CFT and its double
quantization to predict an O(1) quantum  contribution to the partition
functions. Next we analyze the dual picture at the boundary in section 3 in
order to detect this O(1) contribution to the partition function on the CFT side. In
section 4 we write down the spinor holographic formula. In section 5, the pure
AdS bulk geometry is considered as an instance where both sides of the formula
can be worked out in detail. Section 6 is concerned with the Dirac operator at
the boundary and the application of the holographic formula to read off the
universal part in the associated Polyakov formulas, the type-A trace anomaly as
well as the functional determinant on round spheres. In section 7 we examine several
scattered results closely related to our calculations. Concluding remarks
are given in section 8, and conventions and useful identities for Barnes' gammas are 
collected in an appendix.

\section{Bulk spinors and double quantization}

The role of bulk spinors in AdS/CFT has, of course, been extensively studied
since the early days of the correspondence; a non-exhaustive list includes
\cite{Henningson:1998cd,Mueck:1998iz,Henneaux:1998ch,Arutyunov:1998ve}. To begin
with, we choose the bulk side and review the features relevant to our present
concern, namely, the spinor version of the holographic formula.

Consider a bulk metric that approaches asymptotically the Poincar\'e half-space
model for the Euclidean section of $AdS_{n+1}$, that is, hyperbolic space
$H^{n+1}$

\begin{equation}
ds^2= \frac{dz^2+ d\vec{x}^2}{z^2}~.
\end{equation}
The solutions of Dirac equation $(\slash{\!\!\!\nabla}+m )\psi=0$, with positive
mass $m>0$ for definiteness, behave near the conformal boundary $z=0$
as\footnote{Strictly speaking, valid for $0<m<\frac{1}{2}$.}
\begin{equation}
 \psi\sim z^{\lambda_-}\psi_o(\vec{x}) + z^{\lambda_+}\chi_o(\vec{x})~,
\end{equation}
with $\lambda_{\pm}=\frac{n}{2}\pm m$, and the boundary data $\psi_o$ and
$\chi_o$ belong to the eigenspace of the flat Dirac gamma associated to the
$z-$direction, $\Gamma^o$, with eigenvalues $-1$ and $+1$, respectively.

The requirement of regularity at the deep interior, $z\rightarrow\infty$,
imposes a linear relation between $\psi_o$ and $\chi_o$ given by convolution
with the scattering operator $\chi_o=S(\lambda)\ast \psi_0$, or equivalently,
with the kernel associated to the two-point function of the dual operator at
the boundary. It is this on-shell relation which ultimately leads, upon
functional differentiation of the action with respect to the boundary source,
to the corresponding two-point correlator
\begin{equation}
\label{two-flat}
\langle \mathcal{O}_+\;\overline{\mathcal{O}}_+\rangle\sim
\frac{\vec{\Gamma}\cdot(\vec{x}-\vec{y})}{|\vec{x}-\vec{y}|^{1+n+2m}}~.
\end{equation}

The standard AdS/CFT recipe contemplates $\psi_o$ as the  source and $\chi_o$ as
expectation value of the dual primary operator with conformal dimension
$\lambda_+$ at the boundary. In the variational approach to the path integral,
the sum over histories is preformed with $\psi_o$ and $\bar{\psi}_o$ prescribed,
whereas $\chi_o$ and $\bar{\chi}_o$ are free to vary. Nonetheless, a crucial observation 
(cf.~\cite{Allais:2010qq,Klebanov:2011gs,Laia:2011wf})
is that whenever $0<m<1/2$ the dimension $\lambda_-$ is above the unitarity
bound, in this case one is free to interchange the roles of $\psi_o$ and $\chi_o$ to obtain the
two-point function of a dual operator of dimension $\lambda_-$

\begin{equation}
\langle \mathcal{O}_-\;\overline{\mathcal{O}}_-\rangle\sim
\frac{\vec{\Gamma}\cdot(\vec{x}-\vec{y})}{|\vec{x}-\vec{y}|^{1+n-2m}}~.
\end{equation}

\subsection{O(1) contribution to the partition function}

That is very much the picture at the classical level in the bulk. Considering
now quantum fluctuations, we go off-shell and there are
two possible AdS-invariant quantizations of the bulk spinor: the conventional one
with $\psi_o$ set to zero and another, `alternate' one, with $\chi_o$ set
to zero. Whenever the mass of the bulk spinor lies in the window
$0<m<\frac{1}{2}$, both kinds of modes are normalizable (finite energy
configurations~\cite{Laia:2011wf}).
Their contribution to the partition function can be
computed in the standard way via the Green function for the conventional
modes ($\lambda_+$), and analytic continuation to account for the alternate
modes ($\lambda_-$). With this choice at hand, we can emulate the scalar
case~\cite{Breitenlohner:1982jf} since now double quantization for bulk
spinors is established. The relative change in the partition function, upon
functional integration of the quantum fluctuations at quadratic order, is
then given by the ratio of the associated functional determinants:
\begin{equation}
\frac{Z_{grav}^+}{Z_{grav}^-}\;=\;\frac{\det_+
\{\slash{\!\!\!\nabla}_{_X}+m\}}{\det_-\{\slash{\!\!\!\nabla}_{_X}+m\}}~.
\end{equation}

\section{Boundary double-trace deformation}

These two choices of asymptotic behavior correspond in AdS/CFT to two CFT's~\cite{Laia:2011wf,Klebanov:1999tb}
that share the same field content but differ in the dimension of the fermionic
operator $\mathcal{O}$, dual to the bulk spinor. Despite its appearance, the situation is by no means
symmetric; the UV CFT with $\mathcal{O}_-$, perturbed by the {\em relevant} double-trace
deformation $\mathcal{O}^2_-$, flows into the IR CFT with $\mathcal{O}_+$ (cf.~\cite{Allais:2010qq,Klebanov:2011gs,Laia:2011wf}).

\subsection{O(1) contribution to the partition function: a shortcut}

To get a handle on the relative change in the CFT partition functions at the end
points of the RG flow, instead of considering the auxiliary
field trick as in~\cite{Allais:2010qq,Klebanov:2011gs}, we simply  adapt the heuristic
argument given in~\cite{Witten:2003ya} to relate $Z_{UV}$ to $Z_{IR}$.
Namely, in the path integral of the UV CFT we promote the sources $\eta$ and
$\bar{\eta}$ to dynamical fields and integrate over them
\begin{equation}
\frac{Z_{IR}}{Z_{UV}}=\int\,\mathcal{D}\bar{\eta} \,\mathcal{D}\eta
\;\langle\,\exp{\int\,(\bar{\eta}\mathcal{O}_-+\bar{\mathcal{O}}_-\eta)}\,
\rangle ~.
\end{equation}
The expectation value can be approximated, at leading large N due to the
factorization of the correlation functions, by
\begin{equation}
\exp{\int\bar{\eta}\langle
\mathcal{O}_-\;\overline{\mathcal{O}}_-\rangle\eta}~,
\end{equation}
and the Gaussian integral results in the functional determinant of the two-point
function
\begin{equation}
\frac{Z_{IR}}{Z_{UV}}=\det\;\langle
\mathcal{O}_-\;\overline{\mathcal{O}}_-\rangle~,
\end{equation}
or, alternatively,
\begin{equation}
\frac{Z_{UV}}{Z_{IR}}=\det\;\langle
\mathcal{O}_+\;\overline{\mathcal{O}}_+\rangle~.
\end{equation}

\section{The holographic formula}

We have now all necessary ingredients to write down the `spinor holographic
formula' that stems from the postulated equality of the partition functions
in AdS/CFT correspondence, at subleading order O(1):
\begin{equation}
\frac{\det_-
\{\slash{\!\!\!\nabla}_{_X}+m\}}{\det_+\{\slash{\!\!\!\nabla}_{_X}+m\}}=
\det\;\langle
\mathcal{O}_{\lambda}\;\overline{\mathcal{O}}_{\lambda}\rangle_{_\mathcal{M}}
\end{equation}
The `+' means that we compute with the standard $\lambda$ in the asymptotic
behavior of the bulk spinor, whereas `$-$' means the analytic continuation
$\lambda\rightarrow n-\lambda$. As usual, to make sense out of this formula one
needs to tame the divergencies that arise due to the IR divergent  volume of AdS
and the UV divergent short distance singularities. Both divergencies turn out to
be tied by the IR/UV connection in AdS/CFT correspondence~\cite{Susskind:1998dq}.

This formula conjecturally applies to bulk geometries $X$ which are Euclidean
sections of asymptotically locally AdS (ALAdS). In particular, when the
conformal infinity $\mathcal{M}$ belongs to the conformal class of the standard
round spheres, and therefore conformally flat, the bulk is locally AdS and
the IR-divergent volume of AdS factorizes. One can then read off an O(1)
contribution to the holographic trace anomaly in even $n$, just as in the case of
a bulk scalar~\cite{Witten:1998qj,Gra99,HS98}.
Alternatively, from the difference of one-loop effective actions one can compute
the holographic type-A trace anomaly coefficient $a$ following the general recipe
of~\cite{Imbimbo:1999bj}. The behavior of this coefficient for even $n$ and of a
related quantity for odd $n$, in the cases we will explore, gives support to a
conjectured c-theorem valid in all dimensions~\cite{Myers:2010tj} \footnote{This
promises to settle the disparity pointed out in~\cite{Gubser:2002zh} that, although the
computation on the AdS side contemplates even and odd dimensions on equal footing,
it is not clear how to translate the ``holographic'' central charge into field theory
language in the case of odd-dimensional CFT's.} and to an F-theorem
proposal~\cite{Klebanov:2011gs} as well.

\section{The canonical case}

As might be expected, the ball model for hyperbolic space (Euclidean AdS) turns
out to be the simplest bulk background where calculations can be spelled out in
detail and related to the CFT on the conformal boundary (the conformally flat class
of the standard round sphere).

\subsection{Bulk}

The effective action for a Dirac spinor in hyperbolic space has been recently
revisited~\cite{Basar:2009rp,Aros:2010ng} in connection with a curious
gauge-gravity duality where Barnes' multiple gamma function plays a central role.
We briefly survey the relevant steps in the computation of the effective action
\begin{equation}
S^{+}_{grav}=-\log \det\{\slash{\!\!\!\nabla}+m\}~.
\end{equation}
In terms of the Green's function, one has
$(\slash{\!\!\!\nabla}+m)\mathcal{D}=-\mathbb{I}$,
\begin{equation}
S^{+}_{grav}= \int^m \tr \mathcal{D}^{(n+1)}~,
\end{equation}
where also the spinor indices are traced out. There is a subtlety regarding the
dimensionality of the representations of the gamma matrices: for $n$ odd, bulk
and boundary representations share the same dimensionality; whereas for $n$ even,
in order to have a Dirac fermion on the boundary, the dimensionality of the bulk
representation must be doubled~\cite{Laia:2011wf}.

We refer to~\cite{Basar:2009rp,Aros:2010ng} for details
of the implementation of dimensional regularization and the nontrivial role of
the bulk volume. In all, one gets a remarkable result, valid for both even and
odd dimensions, in terms of Barnes' multiple gamma
\begin{equation}
\log
\frac{\det_+
\{\slash{\!\!\!\nabla}+m\}}{\det_-\{\slash{\!\!\!\nabla}+m\}}=-2^{1+\lfloor\frac
{n}{2}\rfloor}\cdot\log
\frac{\Gamma_{n+1}(\frac{n+1}{2}+m)}{\Gamma_{n+1}(\frac{n+1}{2}-m)}~.
\end{equation}
In addition, whenever $n$ is even one can read off the trace anomaly as in the
scalar case~\cite{Diaz:2007an,Diaz:2008hy}.
In the present case one essentially gets the  integral of the spinor
Plancherel\footnote{Difference of the effective Lagrangians. The shorthand
notation involves Pochhammer's symbol
$(x)_n\equiv\frac{\Gamma(x+n)}{\Gamma(x)}$~.}
measure(cf.~\cite{Camporesi:1990xx}) times the volume anomaly
$\mathcal{L}_{n+1}=2(-\pi)^{\frac{n}{2}}/\Gamma(1+\frac{n}{2})$
\begin{equation}
\left[ \frac{2}{(2\pi)^{\frac{n}{2}}}
\int^{m}_0d\mu\;\frac{(\frac{1}{2
}+\mu)_{\frac{n}{2}}\cdot(\frac{1}{2}-\mu)_{\frac{n}{2}}}{(\frac{1}{2})_{\frac{n}{2}}}\right]
\cdot
\mathcal{L}_{n+1}~.
\end{equation}

\subsection{Boundary}

For the round $n$-sphere as conformal boundary, the knowledge of the eigenvalues
of the two-point function
$\langle\mathcal{O}_{\lambda}\;\overline{\mathcal{O}}_{\lambda}
\rangle_{_{S^n}}$ and their degeneracies allows a brute-force computation
of the corresponding functional determinant. The appropriate basis is that of
spinor spherical harmonics, and the eigenvalues\footnote{The eigenvalues can be
also read off from the scattering problem in $H^{n+1}$~\cite{Camporesi:1995fb}.} and
degeneracies have been recently computed in connection with fermionic double-trace
deformations~\cite{Allais:2010qq}
\begin{equation}
\label{eigen}
\mbox{eigenvalues:}\quad \pm \frac{\Gamma(l+n/2+\nu + 1/2)}{\Gamma(l+n/2-\nu +
1/2)}~,
\end{equation}
\begin{equation}
\mbox{degeneracies:}\quad
2^{\lfloor\frac{n}{2}\rfloor}\,\frac{(l+n-1)!}{l!\,(n-1)!}~.
\end{equation}
The formal trace is then assembled\footnote{In dimensional regularization, the
sum over degeneracies of a constant term  vanishes, this is why we do not worry much
about the fact that half of the eigenvalues are negative.} as follows:
\begin{equation}
\fl\log\;\det\;\langle
\mathcal{O}_{\lambda}\;\overline{\mathcal{O}}_{\lambda}\rangle_{_{S^n}}=2^{
1+\lfloor\frac{n}{2}\rfloor}
\sum^{\infty}_{l=0}\frac{(n)_l}{l!}\,\log\frac{\Gamma(l+\frac{n+1}{2}+\nu)}{
\Gamma(l+\frac{n+1}{2}-\nu)}~,
\end{equation}
and it can be worked out within dimensional regularization as
in~\cite{Diaz:2007an,Klebanov:2011gs}
\begin{equation}
\label{pole}
\fl\log\;\det\;\langle
\mathcal{O}_{\lambda}\;\overline{\mathcal{O}}_{\lambda}\rangle_{_{S^n}}=-2^{
1+\lfloor\frac{n}{2}\rfloor}\,
\Gamma(-n)\int^{\nu}_0d\mu\left\{\frac{\Gamma(\frac{n+1}{2}+\mu)}{\Gamma(\frac{
1-n}{2}+\mu)}+ (\mu\rightarrow-\mu)\right\}~.
\end{equation}
This very same regularized answer we have already found in~\cite{Aros:2010ng},
(eqn.5), where Barnes's multiple gamma turned up. The renormalized value can be
written, modulo polynomial and logarithmic terms (we refer again to~\cite{Aros:2010ng}
for further details and eqn.6), as the following quotient of Barnes' multiple gamma
functions
\begin{equation}
\log\;\det\;\langle
\mathcal{O}_{\lambda}\;\overline{\mathcal{O}}_{\lambda}\rangle_{_{S^n}}=2^{
1+\lfloor\frac{n}{2}\rfloor}
\,\log\frac{\Gamma_{n+1}(\frac{n+1}{2}+\nu)}{\Gamma_{n+1}(\frac{n+1}{2}-\nu)}~.
\end{equation}
As explained in~\cite{Aros:2010ng}, the gamma factor in front of (\ref{pole}) is
deceiving: only for $n$ even there is a pole and from the residue one can read off
the (integrated) conformal anomaly under conformal rescaling of the
metric
\begin{equation}
\frac{2^{2+\frac{n}{2}}\,(-1)^{n/2}}{n!}\,\int^{\nu}
_0d\mu\;(\frac{1}{2
}+\mu)_{\frac{n}{2}}\cdot(\frac{1}{2}-\mu)_{\frac{n}{2}}~.
\end{equation}
At this point we have a perfect match between bulk and boundary computations. To
illustrate the usefulness of this result, besides being an explicit corroboration
of the holographic formula, we study a particular value of the spinor mass that
unveils the Dirac operator on the boundary and connects with a vast mathematical
literature on determinants of differential operators on spheres
(cf.~\cite{Bra93}-\cite{Yamasaki}).

\section{Holographic life of Dirac operator}

There are two direct ways to identify the Dirac operator at the boundary.
\begin{itemize}
\item First, consider the action of the Dirac operator on the flat space
two-point function (eqn.~\ref{two-flat})
\begin{equation}
\slash{\!\!\!\nabla}\langle
\mathcal{O}_+(\vec{x})\;\overline{\mathcal{O}}_+(\vec{0})\rangle_{_{R^n}}\sim
\frac{\mathbb{I}}{|\vec{x}|^{n+1+2m}}~.
\end{equation}
Here one can recognize the Laplacian $\nabla^2$ in the limit $m\rightarrow
\frac{1}{2}$, in a distributional sense.
\item Second, by simple inspection of the eigenvalues of the kernel
$\langle
\mathcal{O}_+\;\overline{\mathcal{O}}_+\rangle_{_{S^n}}$ on the $n$-sphere (eqn.\ref{eigen}) in the
same limit
$m\rightarrow \frac{1}{2}$
\begin{equation}
\pm\,(\frac{n}{2}+l)~.
\end{equation}
\end{itemize}

\subsection{Type-A trace anomaly and Polyakov formulas}

We are interested in the universal type-A component of the trace anomaly, the coefficient of the Euler 
density or Pfaffian according to the classification in~\cite{Deser:1993yx}. For conformally related metrics $\hat{g}=e^{2w}g$, 
in the conformally flat class,
Branson (see, e.g., \cite{Bra93}) conjectured the following Polyakov-like formula
\begin{equation}
-\log\frac{\det \hat{\slash{\!\!\!\nabla}}^2}{\det
\slash{\!\!\!\nabla}^2}=c_{_{\slash{\!\!\!\nabla}^2}}^{(n)}\,\int_{\mathcal{M}}
w\left(\hat{Q}_n\,dv_{\hat{g}}\,+\,Q_n\,dv_g\right)+...,
\end{equation}
 where the Pfaffian is traded by Branson's Q-curvature $Q_n$, a  central object in conformal geometry 
 with a much simpler (in fact, linear in $w$) transformation rule under conformal rescalings.

This very same structure we read from the bulk computation, the Q-curvature
terms come from the finite conformal variation of the renomalized
volume~\cite{CQY05,Carlip:2005tz,Aros:2006it} and the overall coefficient
$c_{_{\slash{\!\!\!\nabla}^2}}^{(n)}$ is obtained from the effective Lagrangian
at $m=\frac{1}{2}$
\begin{equation}
c_{_{\slash{\!\!\!\nabla}^2}}^{(n)}=4
\frac{c_{\frac{n}{2}}}{(2\pi)^{\frac{n}{2}}}\,\int^{\frac{1}{2}}_0
d\nu\;\frac{(\frac{1}{2}+\nu)_{_{\frac{n}{2}}}\cdot(\frac{1}{2}-\nu)_{_{\frac{n}
{2}}}}{(\frac{1}{2})_{_{\frac{n}{2}}}}~,
\end{equation}
where $c_k=\frac{(-1)^k}{2^{2k}k!(k-1)!}$. All values reported in~\cite{Bra93}
for
$c_{_{\slash{\!\!\!\nabla}^2}}^{(n)}$ are correctly reproduced by this formula.
Interesting spectral invariants, as zeta function at zero argument and conformal index 
(cf.~\cite{Bra93,Diaz:2008hy}), are easily obtained from this coefficient.  

\subsection{Determinant of iterated Dirac on spheres}

The particular value of the determinant of the scattering operator at the mass
value $\frac{1}{2}$ results in the following remarkable expression in terms of
Barnes' multiple gamma function, after use of recurrence (\ref{recurr}),
\begin{equation}
-\log\det\slash{\!\!\!\nabla}^2\,=\,4\cdot2^{^{[\frac{n}{2}]}}\cdot\log\,
\Gamma_n\,(\frac{n}{2})~.
\end{equation}
Barnes' multiple gamma function is known to occur in functional determinants of Laplacians
on spheres (see, e.g., references in~\cite{Dowker:2003ra}). And yet this compact
expression, which correctly reproduces all zeta-regularized values reported in the 
literature(cf.~\cite{BaerSchopka,Yamasaki}), does not seem to have been noticed until now.

A small digression on a conjecture put forward
by B\"ar and Shopka~\cite{BaerSchopka}: they observed that the numerical values
of these determinants tend to $1$ as the dimension $n$ grows; this was proved
in~\cite{Moeller} not only for the Dirac operator, but also for the Yamabe or
conformal Laplacian on the $n$-sphere. Our result indicates that the above findings
amount to establishing the limiting value of Barnes' gamma $\Gamma_n(n/2)$ as
$n\rightarrow\infty$. Furthermore, we can interpret quite naturally this limiting
value by looking at the bulk side of the holographic formula: the two boundary
conditions coincide as $n$ grows, or alternatively,
$\lambda_{\pm}\rightarrow \frac{n}{2}$ so that both
determinants in the quotient approach one another.
This very same argument would predict the same limiting value for all other
operators on the right side of similar holographic formulas; this is the case for
the determinants of GJMS operators on round spheres~\cite{Diaz:2008hy}. In consequence,
we predict the same limiting value of unity; a prediction that can be probed by the
asymptotic analysis of the explicit results obtained in~\cite{Dowker:2010qy} (see eqn.19 therein).
In fact, the quotient of Barnes's gammas in the limit of large dimension $d$ and fixed
order $k$ tends to $1$, as was the case for the Dirac operator; at the same time, the reported
numerical values for the multiplicative anomaly $M(d,k)$ hint at a vanishing limit value,
so that the exponential of both contributions in eqn.(19) of~\cite{Dowker:2010qy} should
again render $1$ in the limit.

\section{C-theorem proposals and entanglement entropy}

Our computations contain and, at times, generalize several scattered results
that had been independently derived, most of them in the pursuit of extensions
of the C-theorem to higher dimensions and in connection with certain universal
terms in entanglement entropy. We briefly list few of them.

\paragraph{Holographic C-charge at $\mathcal{O}(1)$:}
\begin{itemize}
\item $n=even$, relative change at order $\mathcal{O}(1)$ in Cardy's central
charge computed in~\cite{Allais:2010qq}.
$C_{UV}-C_{IR}>0$ in the mass window $0<m\leq\frac{1}{2}$. Agreement with the
universal {\em log-term} in entanglement entropy~\cite{Myers:2010tj}.\\

We find agreement between bulk and boundary outcomes - tables (1) and (2)
in~\cite{Allais:2010qq}- and our eqns. (14) and (20), respectively. Our
calculation accounts for the overall coefficient as well.

\item $n=odd$, relative change at order $\mathcal{O}(1)$ in one-loop effective
action, candidate for central charge, computed in~\cite{Allais:2010qq}.
$C_{UV}-C_{IR}>0$ in the mass window $0<m\leq\frac{1}{2}$. Agreement with the
universal {\em constant term} in entanglement entropy~\cite{Myers:2010tj}.\\

We find agreement between bulk outcome -integral of eqn.(3.37)
in~\cite{Allais:2010qq}- and our eqn.(13).
We are also able to fill the `hole' left in~\cite{Allais:2010qq} by
identifying this finite contribution on the CFT side, including the overall
coefficient as well (eqn.19).
\end{itemize}

\paragraph{F-coefficient of odd-dimensional CFT's:}

\begin{itemize}
\item F-coefficients for free massless Dirac fermions on odd-spheres, table (2)
in~\cite{Klebanov:2011gs}.\\

This agrees with our result\footnote{The evaluation of the corresponding Barnes' gammas
can be performed with the relations collected in~\ref{AppendixA}, and one can rewrite in
terms of Riemann zeta by use of the relation
$\zeta'(-2n)=\frac{(-1)^n(2n)!}{2^{1+2n}\pi^{2n}}\,\zeta(1+2n)$~.} in eqn.(25) and, of course,
with the values reported in~\cite{BaerSchopka,Moeller}.
The decrease of the numerical values in the above table is precisely the hint
for the conjecture by B\"ar and Schopka.

\item Under RG flow triggered by a fermionic double-trace deformation, at
leading large N, the change in free energy
in three dimensions is given by eqn.(82),\cite{Klebanov:2011gs}.
$F_{UV}-F_{IR}>0$ in the mass window $0<m\leq\frac{1}{2}$~.\\

This coincides again with our bulk (eqn.13) and boundary (eqn.19) results.

\end{itemize}

\paragraph{Entanglement entropy:}

\begin{itemize}
\item Universal terms in entanglement entropy~\cite{Myers:2010tj}: logarithmic
and constant in even and odd dimensions, respectively. In the case of a free
boson, they coincide with the holographic anomaly and determinant of Yamabe 
operator or conformal Laplacian, for even~\cite{Dowker:2010qy} and
odd~\cite{Dowker:2010yj} dimensions, respectively.\\

This matching should hold as well for the entanglement entropy of a free Dirac
spinor and the trace anomaly and determinant of the Dirac operator on $n$-spheres.
Now under the guise of holographic C-charge at $\mathcal{O}(1)$, in the notation 
of~\cite{Myers:2010tj}, they verify $(a^*_d)_{UV}>(a^*_d)_{IR}$. 
\end{itemize}

\section{Conclusion}

Our main contribution has been the spinor version of the holographic formula that
connects the functional determinant of a bulk spinor with that of the two point function of
the dual operator at the boundary. The case of pure AdS allows for explicit computations
which, in turn, encompass several results independently derived in other contexts.
In particular, contact is made in the case of the Dirac operator on the
boundary; here we have obtained a compact expression for the determinant on spheres
in terms of Barnes' gamma function as well as the generic type-A trace anomaly in any
even dimension. 

We have also gained insight into the conjecture by B\"ar and Schopka and
unveiled its possible holographic roots. It seems plausible that the conjecture
should  also apply to the functional determinant of {\em any} conformally covariant 
differential operator, provided a corresponding holographic formula exits.
Natural candidates for the bulk side of such holographic formula are the functional determinants 
in the one-loop effective action of higher-spin fields (see, e.g., recent 
progress in~\cite{Gaberdiel:2010ar}).   

Further study of the holographic formula, for instance in black-hole backgrounds, remains a challenge. 
It seems worth to explore the connection of the functional determinants involved in the formula with 
regularized products of quasinormal frequencies~\cite{Denef:2009kn} or scattering resonances.
Another possible avenue concerns an intriguing feature of Barnes' double gamma at integer values, it 
equals the result of ordinary determinants of Hankel matrices; this might well be a sign for 
{\em localization} of the functional integrals that we have been computing in AdS 
(see, in this respect, \cite{Dabholkar:2011ec}).

We close by noticing that the explicit expressions, as obtained via the holographic approach, 
come out in a gracious, simple form. This was the case for the trace anomaly of GJMS 
operators~\cite{Diaz:2008hy} and now for the determinant and trace anomaly of Dirac operator.

\ack We thank A. Montecinos for collaboration in the initial stages of this study.
This work was partially funded through Fondecyt-Chile 11110430, UNAB DI-21-11/R
and UNAB DI-26-11/R.
\appendix

\section{Barnes' multiple gamma: disambiguation}
\label{AppendixA}

There are several choices for the normalization of Barnes' multiple gamma
function $\Gamma_n(z)$.
For definiteness, we stick to the convention
in~\cite{Yamasaki,Ruijsenaars,FriedmanRuijsenaars} and
list few properties that are relevant to our calculations.

\begin{itemize}
\item Its logarithm can be written in terms of derivatives of Hurwitz zeta
function
\begin{equation}
\log\Gamma_n(z)=\sum^{n-1}_{k=0} b_{n,k}(z)\cdot\zeta'(-k,z)~,
\end{equation}
where $b_{n,k}(z)$ is a polynomial in $z$ with Stirling numbers of the first
kind $s(n,j)$ in its coefficients
\begin{equation}
b_{n,k}(z)=\frac{(-1)^{n-1-k}}{(n-1)!}\,\sum^{n-1}_{j=k}(
\begin{array}{c} j \\ k \\
\end{array})\cdot s(n,j+1)\cdot z^{j-k}~.\end{equation}

\item Recurrence or ladder relation:
\begin{equation}
\label{recurr}
\Gamma_{n+1}(1+z)=\frac{\Gamma_{n+1}(z)}{\Gamma_{n}(z)}~.
\end{equation}
\item Pascal triangle by successive applications of the ladder relation:
\begin{equation}
\fl
\log\Gamma_{n}(m+z)=\sum^{m}_{l=0}(-1)^l \,(
\begin{array}{c} m \\ l \\
\end{array})\cdot\log\Gamma_{n-l}(z)\,,\qquad\quad 0\leq m\leq n-1~.
\end{equation}
\item Particular values in terms of derivatives of Riemann zeta
function\footnote{Relation \ref{Gamma_n_at_1} can be further simplified to
$\log\Gamma_n(1)=\frac{1}{(n-1)!}\sum^{n-1}_{j=1} |s(j,n-1)|\cdot\zeta'(-k)$,
cf.~\cite{QuineChoi} where their $P_n$ is just
the inverse of the gamma we are using.}:
\begin{equation}
\label{Gamma_n_at_1}
\log\Gamma_n(1)=\sum^{n-1}_{k=0} b_{n,k}(1)\cdot\zeta'(-k)~,
\end{equation}

\begin{equation}
\fl
\log\Gamma_n(\frac{1}{2})=\sum^{n-1}_{k=0}
b_{n,k}(\frac{1}{2})\cdot(2^{-k}-1)\cdot\zeta'(-k)
-\log2\sum^{n-1}_{k=0} b_{n,k}(\frac{1}{2})\frac{2^{-k}B_{k+1}}{k+1}~,
\end{equation}
where $B_n$ are Bernoulli numbers.\\

\end{itemize}

\vspace{0.5cm}

\section*{References}

\providecommand{\href}[2]{#2}\begingroup\raggedright\endgroup
\end{document}